\documentclass{ws-ijmpcs}

\usepackage{amsmath,epsfig,axodraw}
\usepackage{graphicx}
\usepackage[figuresright]{rotating}

\newcommand{\amu}{a_\mu}
\newcommand{\amuSUOL}{a_\mu^{\rm 1L}}

\newcommand{\amub}{a_{\mu;\text{bos}}^{\rm EW(2)}}

\newcommand{\amufrestH}{a_{\mu;\text{f-rest,H}}^{\rm EW(2)}}

\newcommand{\amufrestnoH}{a_{\mu;\text{f-rest,no H}}^{\rm EW(2)}}
\begin{document}

\markboth{H.\ G.\ Fargnoli, C.\ Gnendiger, S.\ Pa{\normalfont\ss}ehr,
  D.\ St\"ockinger, H.\ St\"ockinger-Kim}
{Electroweak Standard Model
  prediction for $(g-2)_\mu$ and improvements on the MSSM prediction}

\catchline{}{}{}{}{}
%
%%%%%%%%%%%%%%%%%%%%%%%%%%%%%%%%%%%%%%%%%%%%%%%%%%%%%%%%%%%%%%%%%%%%

% declarations for front matter
\title{The full electroweak Standard Model
  prediction for $(g-2)$ of the muon and improvements on the MSSM
  prediction\footnote{Contribution to the Proceedings for the International Workshop on
    $e^+e^-$ collisions from Phi to Psi, Rome, September 2013.}}

\author{H.\ G.\ Fargnoli$^a$, C.\ Gnendiger$^b$, S.\ Pa{\normalfont\ss}ehr$^c$,
  D.\ St\"ockinger$^b$, H.\ St\"ockinger-Kim$^b$}
\address{$^a$ Universidade Federal de Lavras, Lavras, Brazil\\
$^b$ Institut f\"ur Kern- und Teilchenphysik,
TU Dresden, Dresden, Germany\\
$^c$ Max-Planck Institut f\"ur Physik,
M\"unchen, Germany}

\maketitle

\begin{abstract}
Recent progress on the $(g-2)_\mu$ prediction is
presented. In the SM, the Higgs-boson mass dependent contributions have been evaluated
exactly up to the two-loop level and consistently combined with
leading three-loop effects. Thus, the currently most accurate
value including a detailed error analysis for the SM electroweak
contributions has been obtained. The SUSY two-loop 
corrections from fermion/sfermion-loop insertions have been computed;
they are generally large and logarithmically enhanced for heavy squarks.
\end{abstract}

\section{Introduction}

The
deviation between the measurement\cite{Bennett:2006} of the
muon anomalous magnetic moment $a_\mu=(g-2)_\mu/2$ and 
the SM prediction\cite{SMreviews}
%(see Refs.\
%\cite{Gnendiger:2013pva,Davier,HMNT,Benayoun:2012wc,MdRRS} for 
%recent evaluations and reviews)
 amounts to more than $3\sigma$. It is
 of utmost interest to further scrutinize this longstanding deviation.
% Indeed at this conference progress on all fronts has been
% discussed---on the experimental
% determination\cite{Fermilabproc,JParcproc}, the QED
% theory\cite{Steinhauserproc} and on all aspects of the hadronic
% contributions\cite{hadproc}.

In these proceedings we discuss a recent
reevaluation\cite{Gnendiger:2013pva} of the SM electroweak
contributions. These 
are the only SM contributions with a 
dependence on the Higgs boson mass. Earlier evaluations had an
irreducible theory uncertainty since the Higgs boson mass
used to be unknown. After the Higgs boson mass has been measured at the LHC, we are now in
the position to obtain the final value of the Higgs-dependent part of
the SM prediction for $a_\mu$ (see Sec.\ \ref{sec:EW}). 

We also briefly comment on complementary progress\cite{shortfsfloops,longfsfloops}
for the prediction of $a_\mu$ in the minimal supersymmetric standard
model (MSSM) (see Sec.\
\ref{sec:SUSY}).

\section{The electroweak contributions after the Higgs boson mass
  measurement}
\label{sec:EW}

Before the Higgs boson mass measurement, the most precise
evaluation\cite{CzMV} of the electroweak contributions obtained the
result 
$a_\mu^{\rm EW}=(154\pm1\pm2)\times10^{-11}$,
where the first error is due to electroweak hadronic uncertainties,
but the second, larger uncertainty is due to the unknown Higgs boson mass.

\begin{figure}
\null\vspace{-2.5em}
\includegraphics[width=0.45\textwidth]{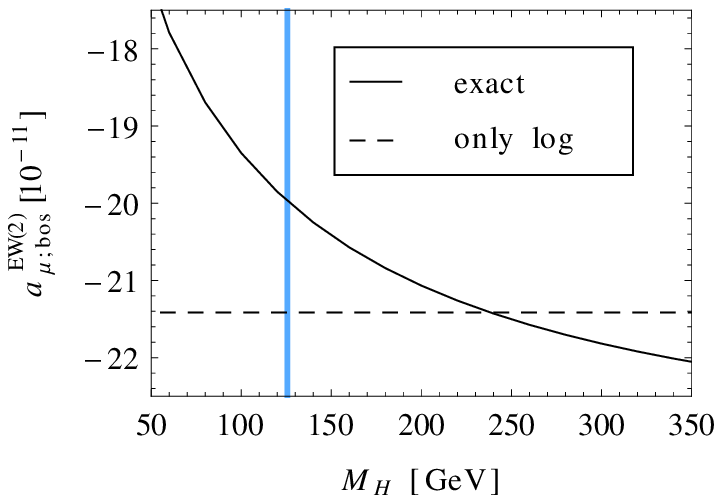}
\includegraphics[width=0.45\textwidth]{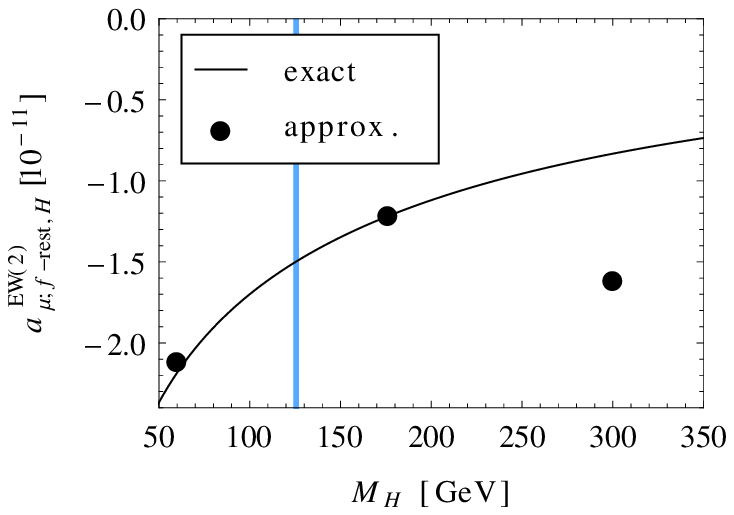}
\vspace{-2em}
\caption{
Numerical result for $\amub$ (left) and  for $\amufrestH$ (right)
  as a function of the Higgs boson mass. The vertical
  band indicates the measured value of $M_H$. The dashed line
  in the left plot corresponds to the leading logarithmic
  approximation as defined in Ref.\ \protect\cite{HSW0304}. The fat dots
  in the right plot correspond to the approximations for
  $M_H=60\ \mbox{GeV},m_t,300\ \mbox{GeV}$ given in Ref.\ \protect\cite{CKM1}. 
\label{fig:Higgsplots}
}
\end{figure}

In the update\cite{Gnendiger:2013pva} of this result we take into
account the following aspects:
\begin{itemize}
\item All Higgs-boson mass dependent contributions up to the two-loop
  level are exactly evaluated, using the value\cite{ATLASCMS}
$M_H=125.6$~GeV, with a conservative error of $\pm1.5\ {\rm GeV}$. As further input
parameters, the masses of muon, Z-boson and top quark,
the muon decay constant $G_F$ and the fine-structure constant $\alpha$
in the Thomson limit, are chosen\cite{PDG2012}.
\item The W-boson mass is then unambiguously predicted by the
SM\cite{Awramik:2003rn}, and we use the appropriate theory value
$M_W = 80.363 \pm 0.013\ \mbox{GeV}$.
\item An analysis of the theory uncertainty, in particular due to the
  uncertainty of the SM input parameters, is carried out.
\item The exact evaluation of the Higgs-dependent contributions is
  consistently combined with the remaining contributions up to the
  three-loop level. In order to avoid double counting, the choice of $\alpha$ in the
  two-loop contributions has to match the one in the evaluation of
  leading three-loop contributions. We choose the Thomson-limit definition
  for $\alpha$ at the two-loop level; see Eq.\ (\ref{EWthreeloop})
  below for the consequences. 
\end{itemize}
The most important new results are the exact results of the 
Higgs-dependent two-loop contributions.
Fig.\ \ref{fig:Higgsplots} shows them for a range of Higgs
boson masses. The left panel shows the
bosonic contributions $\amub$ (i.e.\ without fermion loop), the right panel
the Higgs-dependent fermionic contributions $\amufrestH$. For the
input parameters given above, these contributions, and the one-loop
contribution amount to
\begin{align}
a_\mu^{\rm EW(1)}
&=(\,194.80\pm 0.01)\times 10^{-11},
\label{oneloop}
\\
\label{Higgsbosonicres}
\amub&=(-19.97\pm 0.03)\times 10^{-11},
\\
\amufrestH&=(\,\ -1.50\pm0.01)\times10^{-11}.
\label{Higgsfermionicres}
\end{align}
The results are exact up to the parametric uncertainties due
to the uncertainty of the input parameters $M_W$, $m_t$, and $M_H$.
The dominant uncertainty arises in $\amub$ due to the uncertainty of
the Higgs boson mass. Nevertheless, the overall uncertainty of these
contributions is of the order $10^{-13}$ and thus extremely tiny. 

We briefly comment on the
difference between these results and earlier results in the literature.

The first computation\cite{CKM2} of $\amub$ was a milestone but 
  employed an approximation assuming $M_H\gg M_W$. Refs.\ 
\cite{HSW0304,CzarneckiGribouk} provided the full
$M_H$-dependence of $\amub$, however only in (semi)numerical form and for a
particular, different set of input parameters than the one used here.

The first computation of the Higgs-dependent fermionic contributions
$\amufrestH$ was
  carried out in Ref.\ \cite{CKM1} in three limiting cases, \mbox{$M_H\ll m_t$},
\mbox{$M_H= m_t$}, \mbox{$M_H\gg m_t$}. Furthermore, the approximation
$s_W^2=1/4$ was used and diagrams with $\gamma$--$Z$--Higgs subdiagram were
neglected.
The results of Fig. \ref{fig:Higgsplots} (right)
show that the approximation of Ref.\ \cite{CKM1} for large $M_H$ is 
surprisingly poor, compared to the exact result.  There, the
higher-order terms in $m_t^2/M_H^2$ are 
important. 

The remaining two-loop contributions beyond $\amub$ and $\amufrestH$ are non-Higgs dependent
  fermion-loop contributions: $\amu^{\rm EW(2)}(f)$, the diagrams with
  $\gamma\gamma Z$ interaction generated by a fermion $f$-loop;
  $\amufrestnoH$, the remaining two-loop contributions from fermion
  loops. These have been evaluated in Ref.\ \cite{CzMV}, after 
earlier work\cite{CKM1,PerisKnecht}
particularly on  $\amu^{\rm EW(2)}(f)$. 
 The leading three-loop contributions enhanced by large
  logarithms have been evaluated in
Refs.\ \cite{DG98,CzMV}.

The results for these remaining, non-Higgs dependent electroweak
two-loop and leading three-loop contributions, are\cite{CzMV}
\begin{align}
\label{amuEWf}
a_\mu^{\rm EW(2)}(e,\mu,u,c,d,s)&=-(6.91\pm0.20\pm0.30)\times10^{-11},
\\
a_\mu^{\rm EW(2)}(\tau,t,b)&=-(8.21\pm0.10)\times10^{-11},
\\
\amufrestnoH&=-(4.64\pm0.10)\times10^{-11},\\
a_\mu^{\rm EW(\ge3)}&=(0\pm0.20)\times10^{-11}.
\label{EWthreeloop}
\end{align}
Eq.\ (\ref{EWthreeloop}) is correct for the parametrization of the two-loop
result in terms of $\alpha$, while the result for the alternative
parametrization in terms of $G_F$ would have been\cite{CzMV} $a_\mu^{\rm
    EW(\ge3)}=(0.40\pm0.20)\times10^{-11}$. 
The dominant theory error arises in  the first two generations in $a_\mu^{\rm
  EW(2)}(e,\mu,u,c,d,s)$. It has been estimated\cite{CzMV} by varying
the hadronic input parameters and by estimating higher-order QCD
corrections.
% As noted above and discussed in Ref.\ \cite{CzMV} the
%three-loop corrections accidentally cancel, if the two-loop
%contributions are parametrized in terms of $\alpha$ in the Thomson
%limit.

The final result for the electroweak SM contributions to $a_\mu$ is
the sum of all presented contributions,
\begin{align}
a_\mu^{\rm EW}&= (153.6\pm1.0)\times10^{-11}.
\label{amuEWNew}
\end{align}
The final theory error of these contributions dominated by the
electroweak hadronic and three-loop uncertainties of 
Eqs.\ (\ref{amuEWf}--\ref{EWthreeloop}). It is enlarged to the
conservative value 
$\pm1.0\times10^{-11}$, in line with Ref.\ \cite{CzMV}. 
 The parametric uncertainty
due to the input parameters $M_W$, $m_t$, and particularly $M_H$ is
negligible. The precision
of the result is by far sufficient for the next generation of $a_\mu$
measurements. Clearly, the full Standard Model theory error remains dominated by the
non-electroweak hadronic contributions. 

\section{Non-decoupling two-loop contributions in the MSSM}
\label{sec:SUSY}

Supersymmetry (SUSY) is a promising explanation of
the $3\sigma$ deviation  in $a_\mu$, although simple
SUSY scenarios where all SUSY particles are  light are already
ruled out by the LHC.  
Ref.\ \cite{shortfsfloops} has defined several
benchmark parameter points which illustrate that already the one-loop SUSY contributions
to $a_\mu$ have an intricate parameter dependence, if non-trivial SUSY
mass patterns are allowed:  SUSY contributions in the ballpark of
the current $3\sigma$ deviation can be obtained if, e.g., the Higgsino mass $\mu$ is
much larger than the bino mass $M_1$ and the smuon masses, or if the
wino mass $M_2$ and the left-handed smuon mass are much larger than
$\mu$, $M_1$ and the right-handed smuon mass. Parameter scenarios with
large $\mu$ have been studied including leading higher-order
corrections also in   Refs.~\cite{Endo} recently.

\begin{figure}
\null\vspace{-2.5em}
\begin{center}
\scalebox{.65}{\setlength{\unitlength}{1pt}
\begin{picture}(240,100)(-120,-20)
\CArc(0,0)(60,0,60)
\CArc(0,0)(60,120,180)
\ArrowArcn(0,51)(30,180,0)
\DashArrowArcn(0,51)(30,0,180){4}
\ArrowLine(-120,0)(-60,0)
\ArrowLine(60,0)(120,0)
\DashArrowLine(-60,0)(60,0){4}
\Photon(-100,100)(-60,60){4}{5}
\Vertex(30,51){2}
\Vertex(-30,51){2}
\Vertex(60,0){2}
\Vertex(-60,0){2}
\Text(-80,95)[]{\scalebox{1.35}{$\gamma$}}
\Text(-100,-10)[]{\scalebox{1.35}{$\mu$}}
\Text(100,-10)[]{\scalebox{1.35}{${\mu}$}}
\Text(0,-10)[c]{\scalebox{1.35}{$\tilde{\mu}, \tilde{\nu}_{\mu}$}}
\Text(-40,20)[]{\scalebox{1.35}{$\tilde{\chi} _{j} ^{0,-}$}}
\Text(40,20)[]{\scalebox{1.35}{$\tilde{\chi} _{i} ^{0,-}$}}
\Text(0,70)[]{\scalebox{1.35}{$f,f'$}}
\Text(0,32)[]{\scalebox{1.35}{$\tilde f$}}
\end{picture}}
\scalebox{.95}[.9]{\begin{picture}(190,125)
\put(37,25){\epsfxsize=0.17\textwidth\epsfbox{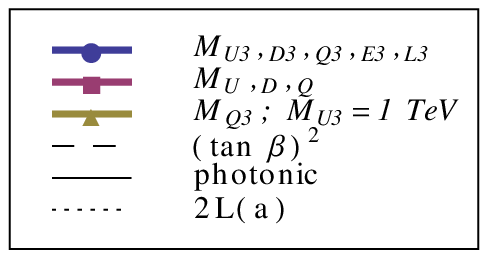}}
\put(10,-10){\epsfxsize=0.4\textwidth\epsfbox{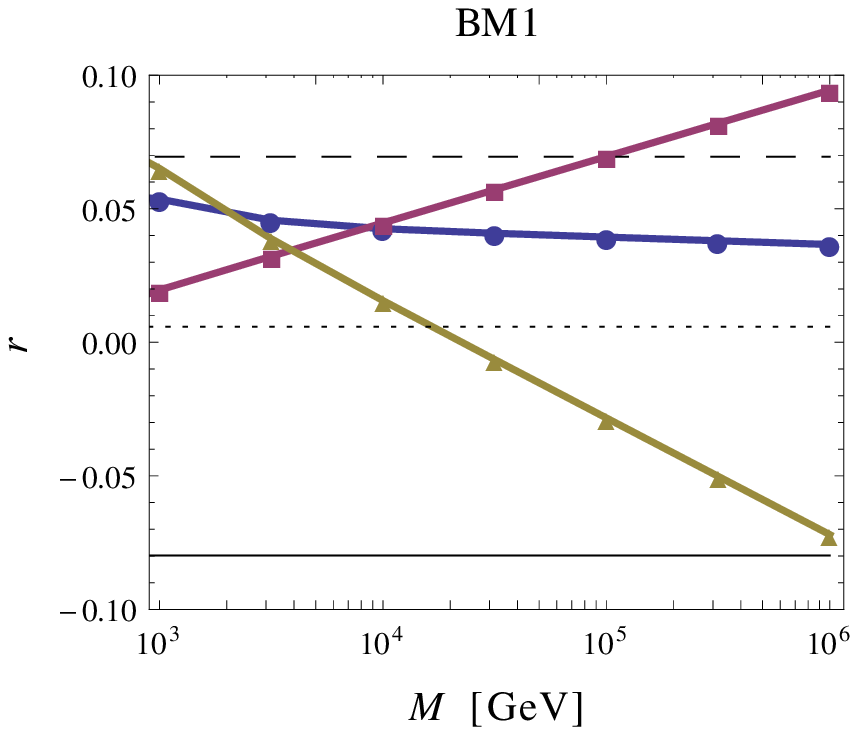}}
\end{picture}}
\end{center}
\caption{\label{fig:diagrams}
Left: SUSY two-loop diagrams with
  fermion/sfermion-loop insertion. Right: Results for the ratios  $r\equiv{a_\mu^{\rm 2L}}/{\amuSUOL}$
  for various classes of MSSM two-loop contributions $a_\mu^{\rm 2L}$
  and the MSSM one-loop contributions $\amuSUOL$. The benchmark point
  BM1 \protect\cite{shortfsfloops} is defined by $\mu=350$ GeV, $M_2=2M_1=300$
  GeV, 
  $\tan\beta=40$. The left- and right-handed smuon masses are 400 GeV.  The thick, coloured lines show 
  the new fermion/sfermion-loop contributions for the combinations
of sfermion masses indicated in the legend. The
thin lines show the previously known $(\tan\beta)^2$ (dashed),
photonic (solid), and 2L(a) (dotted) contributions. }
%%%%%%%%%%%%%%%%%
\end{figure}

Recently, a class of contributions has been computed\cite{shortfsfloops,longfsfloops}
which can become particularly 
important for such split spectra: two-loop contributions where a
fermion/sfermion loop is inserted into a SUSY one-loop diagram, see
Fig.\ \ref{fig:diagrams} (left).

The most prominent features of these two-loop contributions (plus the
associated counterterm contributions) are that
\begin{itemize}
\item[$(i)$] they contain the large universal quantities $\Delta\alpha$ and
$\Delta\rho$ from fermion loops, which make the contributions
generically sizeable;
\item[$(ii)$] they are logarithmically enhanced
if the sfermion masses in the inner loop become large. 
\end{itemize}

These points allow for a very compact
approximation formula\cite{longfsfloops,code}.
Fig.\ \ref{fig:diagrams} (right) illustrates the non-decoupling,
logarithmic enhancement of the contributions for large sfermion
masses. We restrict ourselves to various motivated combinations of 
sfermion soft SUSY breaking parameters $M_{Qi}$, $M_{Ui}$, $M_{Di}$,
$M_{Li}$, $M_{Ei}$ (where the first index denotes the supermultiplet,
the second the generation): either of a common 
third generation sfermion mass $M_{U3,D3,Q3,E3,L3}\equiv M$, or of a
universal first and second generation squark mass
$M_{U1,D1,Q1,U2,D2,Q2}\equiv M$, or, as an example with
particularly 
large corrections, purely as a function of
$M_{Q3}$ with $M_{U3}$ fixed to 1~TeV. Each time, the non-varied
sfermion masses are kept at standard values, which are 7~TeV for
the squark masses and 3~TeV for the third generation slepton
masses. The selectron masses are set to the smuon masses, and the
trilinear $A$ parameters are set to zero.

As the figure shows, these new
fermion/sfermion-loop contributions can be the 
largest MSSM two-loop contributions to $a_\mu$ --- already for
moderate inner sfermion masses. For different sets of input
parameters than shown in the figure, up to 10\%
corrections for small and up to 30\% corrections for large sfermion
masses can be found\cite{shortfsfloops,longfsfloops}. Fig.\
\ref{fig:diagrams} also shows the other known two-loop
contributions: the SUSY corrections to SM-one-loop diagrams (``class
2L(a)'')\cite{HSW0304}, the photonic corrections\cite{vonWeitershausen:2010zr},
and the $(\tan\beta)^2$-corrections\cite{Marchetti:2008hw}, which are
in the range $(-7,\ldots,+7)\%$.

\end{document}